\documentclass{article}
\usepackage{spconf,amsmath,amssymb,booktabs,color,graphicx,hyperref,tabularx,url}
\hypersetup{
    colorlinks=true,
    citecolor=green,
    linkcolor=red,
    urlcolor=magenta,
}
\makeatletter
\newcommand\footnoteref[1]{\protected@xdef\@thefnmark{\ref{#1}}\@footnotemark}
\makeatother
\title{Wave-U-Net Discriminator: Fast and Lightweight Discriminator\\for Generative Adversarial Network-Based Speech Synthesis}

\name{Takuhiro Kaneko, Hirokazu Kameoka, Kou Tanaka, Shogo Seki}
\address{NTT Communication Science Laboratories, NTT Corporation, Japan}

\begin{document}
\ninept

\maketitle

\begin{abstract}
  In speech synthesis, a generative adversarial network (GAN), training a generator (speech synthesizer) and a discriminator in a min-max game, is widely used to improve speech quality. An ensemble of discriminators is commonly used in recent neural vocoders (e.g., HiFi-GAN) and end-to-end text-to-speech (TTS) systems (e.g., VITS) to scrutinize waveforms from multiple perspectives. Such discriminators allow synthesized speech to adequately approach real speech; however, they require an increase in the model size and computation time according to the increase in the number of discriminators. Alternatively, this study proposes a Wave-U-Net discriminator, which is a single but expressive discriminator with Wave-U-Net architecture. This discriminator is unique; it can assess a waveform in a sample-wise manner with the same resolution as the input signal, while extracting multilevel features via an encoder and decoder with skip connections. This architecture provides a generator with sufficiently rich information for the synthesized speech to be closely matched to the real speech. During the experiments, the proposed ideas were applied to a representative neural vocoder (HiFi-GAN) and an end-to-end TTS system (VITS). The results demonstrate that the proposed models can achieve comparable speech quality with a 2.31 times faster and 14.5 times more lightweight discriminator when used in HiFi-GAN and a 1.90 times faster and 9.62 times more lightweight discriminator when used in VITS.\footnote{\label{foot:samples}Audio samples are available at \url{https://www.kecl.ntt.co.jp/people/kaneko.takuhiro/projects/waveunetd/}.}
\end{abstract}


\section{Introduction}
\label{sec:introduction}

Speech plays an important role in human--human and human--machine communications. To obtain the desired or necessary speech for enriching communications, speech synthesis (e.g., text-to-speech (TTS), a technology for producing speech from text) has been actively studied.

One of the most successful approaches for speech synthesis is the two-stage approach. The first model predicts the intermediate representation (e.g., mel spectrogram) from the input data (e.g., text), and the second model synthesizes speech from a predicted intermediate representation. The second model is called a neural vocoder, and various neural vocoders (e.g., autoregressive models~\cite{AOordArXiv2016,NKalchbrennerICML2018}, flow models~\cite{AOordICML2018,RPrengerICASSP2019}, generative adversarial network (GAN) models~\cite{KKumarNeurIPS2019,RYamamotoICASSP2020,JYangIS2020,JKongNeurIPS2020,GYangSLT2021,AMustafaICASSP2021,TKanekoICASSP2022,TKanekoIS2022}, and diffusion probabilistic models~\cite{NChenICLR2021,ZKongICLR2021}) have been proposed. The merit of this approach is the portability of each model, and a learned neural vocoder is commonly used in other tasks, such as voice conversion (e.g.~\cite{KQianICML2019,TKanekoIS2020,TKanekoICASSP2021}).

Another successful approach is the end-to-end approach (e.g.,~\cite{WPingICLR2019,YRenICLR2021,JDonahueICLR2021,JKimICML2021}) that directly converts input data (e.g., text) to speech using a unified model. This approach is advantageous for reducing the cascaded errors caused by connecting two separately trained models and eliminating the bias caused by using nonoptimal intermediate representation.

In both approaches, the common objective is to obtain high-quality speech. Thus, a GAN~\cite{IGoodfellowNIPS2014}, a framework that trains a generator (speech synthesizer) and discriminator in a two-player min-max game, has gained attention and has been widely used both in two-stage~\cite{KKumarNeurIPS2019,RYamamotoICASSP2020,JYangIS2020,JKongNeurIPS2020,GYangSLT2021,AMustafaICASSP2021,TKanekoICASSP2022,TKanekoIS2022} and end-to-end~\cite{YRenICLR2021,JDonahueICLR2021,JKimICML2021} approaches. In particular, an ensemble of discriminators is useful for assessing a waveform from multiple perspectives and is commonly used in recent neural vocoders (e.g., HiFi-GAN~\cite{JKongNeurIPS2020}) and end-to-end TTS systems (e.g., VITS~\cite{JKimICML2021}). Such discriminators succeed in bringing the synthesized speech adequately close to real speech; however, they require an increase in the model size and computation time according to the increase in the number of discriminators.

\begin{figure}[t]
  \centerline{\includegraphics[width=\columnwidth]{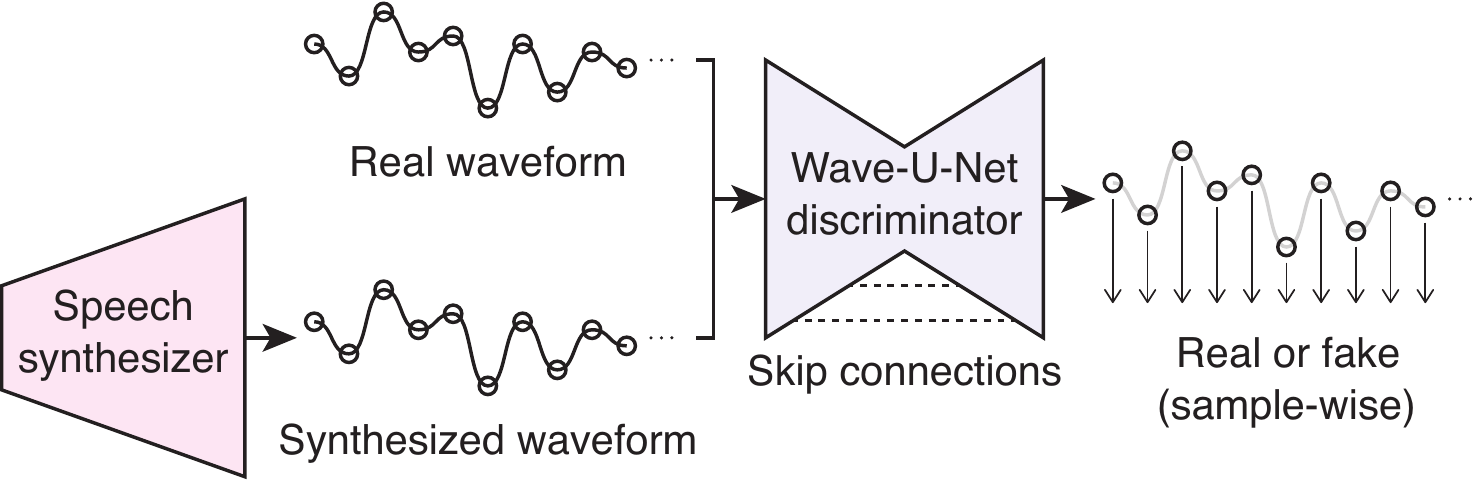}}
  \caption{Overview of GAN training with a Wave-U-Net discriminator. A Wave-U-Net discriminator is unique in that it assesses a waveform in a sample-wise manner in the same resolution as the input signal while extracting multilevel features via an encoder and decoder with skip connections.}
  \label{fig:concept}
\end{figure}

This fact motivates the need to address the following question: \textit{``Can one replace an ensemble of discriminators with a single expressive discriminator?''} As an answer, this study proposes a \textit{Wave-U-Net discriminator}, a novel discriminator with Wave-U-Net architecture~\cite{SPascualIS2017,DStollerISMIR2018}, inspired by the recent success of a U-Net discriminator~\cite{ESchonfeldCVPR2020} (discriminator with a U-Net architecture~\cite{ORonnebergerMICCAI2015}) in image synthesis. Figure~\ref{fig:concept} presents an overview of GAN training with a Wave-U-Net discriminator. As shown in Figure~\ref{fig:concept}, the Wave-U-Net discriminator is unique in that it assesses a waveform in a sample-wise manner with the same resolution as the input signal while extracting multilevel features using an encoder and decoder with skip connections. This architecture provides a generator (speech synthesizer) with sufficiently rich information for synthesized speech to approach real speech. This enabled replacing a typical ensemble of discriminators with a single discriminator.

During the experiments, the general validity of the proposed ideas was investigated by evaluating Wave-U-Net discriminators in various situations: (1) evaluation on neural vocoders with HiFi-GAN~\cite{JKongNeurIPS2020} in various datasets, including LJSpeech~\cite{ljspeech17} (single English speaker), VCTK~\cite{JYamagishiCSTR2016} (multiple English speakers), and JSUT~\cite{RSonobeArXiv2017} (single Japanese speaker), and (2) evaluation on end-to-end TTS with VITS~\cite{JKimICML2021}. The results demonstrate that the proposed model can achieve comparable speech quality with a $2.31$ times faster and $14.5$ times more lightweight discriminator when used in HiFi-GAN and a $1.90$ times faster and $9.62$ times more lightweight discriminator when used in VITS.

The remainder of this paper is organized as follows: Section~\ref{sec:gan} briefly reviews GAN-based speech synthesis. Section~\ref{sec:waveunetd} describes the proposed Wave-U-Net discriminator. Section~\ref{sec:experiments} presents the experimental results, and Section~\ref{sec:conclusion} concludes the paper with remarks on future research.

\section{GAN-based speech synthesis}
\label{sec:gan}

\subsection{Overview}
\label{subsec:gan_overview}

A GAN~\cite{IGoodfellowNIPS2014} is a framework that trains a generator $G$ and discriminator $D$ in a two-player min-max game. In the context of speech synthesis, a generator is identical to a speech synthesizer that synthesizes a speech waveform $x$ from the input data $s$, where $s$ indicates the intermediate representation (e.g., mel spectrogram) in a two-stage approach and text in an end-to-end TTS system. In GAN-based speech synthesis, the generator and discriminator are trained with two GAN-related losses: adversarial and feature-matching losses.

\subsection{Losses}
\label{subsec:gan_losses}

\smallskip\noindent\textbf{Adversarial losses.}
Adversarial losses, particularly the variants of a least squares GAN~\cite{XMaoICCV2017} that are commonly used in speech synthesis, are defined as follows:
\begin{flalign}
  \label{eq:adv_loss_d}
  \mathcal{L}_{Adv}(D) & = \mathbb{E}_{(x, s)}[ (D(x) - 1)^2 + (D(G(s)))^2 ],
  \\
  \label{eq:adv_loss_g}
  \mathcal{L}_{Adv}(G) & = \mathbb{E}_{s}[ (D(G(s)) - 1)^2 ],
\end{flalign}
where $D$ attempts to distinguish between real and synthesized speech by minimizing $\mathcal{L}_{Adv}(D)$ and $G$ attempts to synthesize speech indistinguishable by $D$ by minimizing $\mathcal{L}_{Adv}(G)$.

\smallskip\noindent\textbf{Feature-matching loss.}
To stabilize the adversarial training, a feature-matching loss~\cite{ALarsenICML2016,TKanekoIS2017} is simultaneously used:
\begin{flalign}
  \label{eq:fm_loss}
  \mathcal{L}_{FM}(G) = \mathbb{E}_{(x, s)}\left[ \sum\nolimits_{i = 1}^T \frac{1}{N_i} \| D^i(x) - D^i(G(s)) \|_1 \right],
\end{flalign}
where $T$ indicates the number of layers in $D$, $D^i$ and $N_i$ denote the features and number of features in the $i$-th layer of $D$. Here, $G$ attempts to synthesize speech close to the ground-truth speech by minimizing $\mathcal{L}_{FM}(G)$.

To further stabilize the training, losses other than GAN-related ones were also used in typical GAN-based speech synthesis. For example, reconstruction losses, such as spectrogram loss~\cite{RYamamotoICASSP2020} and mel-spectrogram loss~\cite{JKongNeurIPS2020}, are widely used. Owing to space limitations, this study omits explanations of them. Please refer to the corresponding studies for more details.

\begin{figure}[t]
  \centerline{\includegraphics[width=0.77\columnwidth]{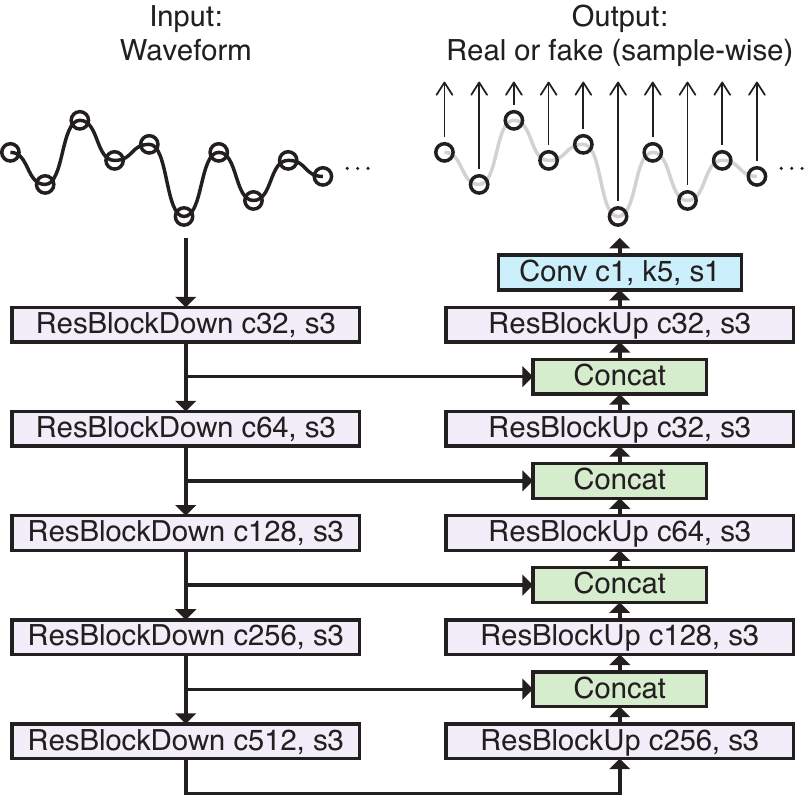}}
  \caption{Overview architectures of a Wave-U-Net discriminator. \texttt{c}$x$, \texttt{k}$y$, and \texttt{s}$z$ indicate the number of output channels of $x$, kernel size of $y$, and stride of $z$, respectively. The details of residual blocks for downsampling (\texttt{ResBlockDown}) and upsampling (\texttt{ResBlockUp}) are presented in Figure~\ref{fig:resblocks}.}
  \label{fig:architecture}
\end{figure}

\section{Wave-U-Net discriminator}
\label{sec:waveunetd}

\subsection{Overview}
\label{subsec:waveunetd_overview}

As discussed in Section~\ref{sec:gan}, in a GAN, the generator is trained and guided by an adversarial discriminator. To guide the learning of the generator in the correct direction, the discriminator must capture a speech waveform sufficiently and propagate adequate information to the generator. In particular, speech waveforms have multilevel (e.g., multiscale and multiperiod) structures; therefore, the discriminator must adequately capture such structures.

To satisfy this requirement with a single discriminator, this study introduced a Wave-U-Net discriminator with Wave-U-Net architecture~\cite{SPascualIS2017,DStollerISMIR2018}. Figure~\ref{fig:architecture} presents an overview of the Wave-U-Net discriminator architecture. As shown in this figure, a Wave-U-Net discriminator differs from a typical discriminator in that a typical discriminator only constitutes an encoder and judges the reality of speech using abstracted features obtained via downsampling. By contrast, a Wave-U-Net discriminator has an encoder-decoder architecture and assesses a waveform in a sample-wise manner with the same resolution as the input waveform by applying upsampling after downsampling. In this process, multilevel features are efficiently extracted using an encoder and decoder with skip connections. This study suggests that the architectural design provides a rich information source for synthesized speech to approach real speech, similar to a typical ensemble of discriminators. We empirically demonstrate the validity of this statement in various situations through the experiments described in Section~\ref{sec:experiments}.

\begin{figure}[t]
  \centerline{\includegraphics[width=\columnwidth]{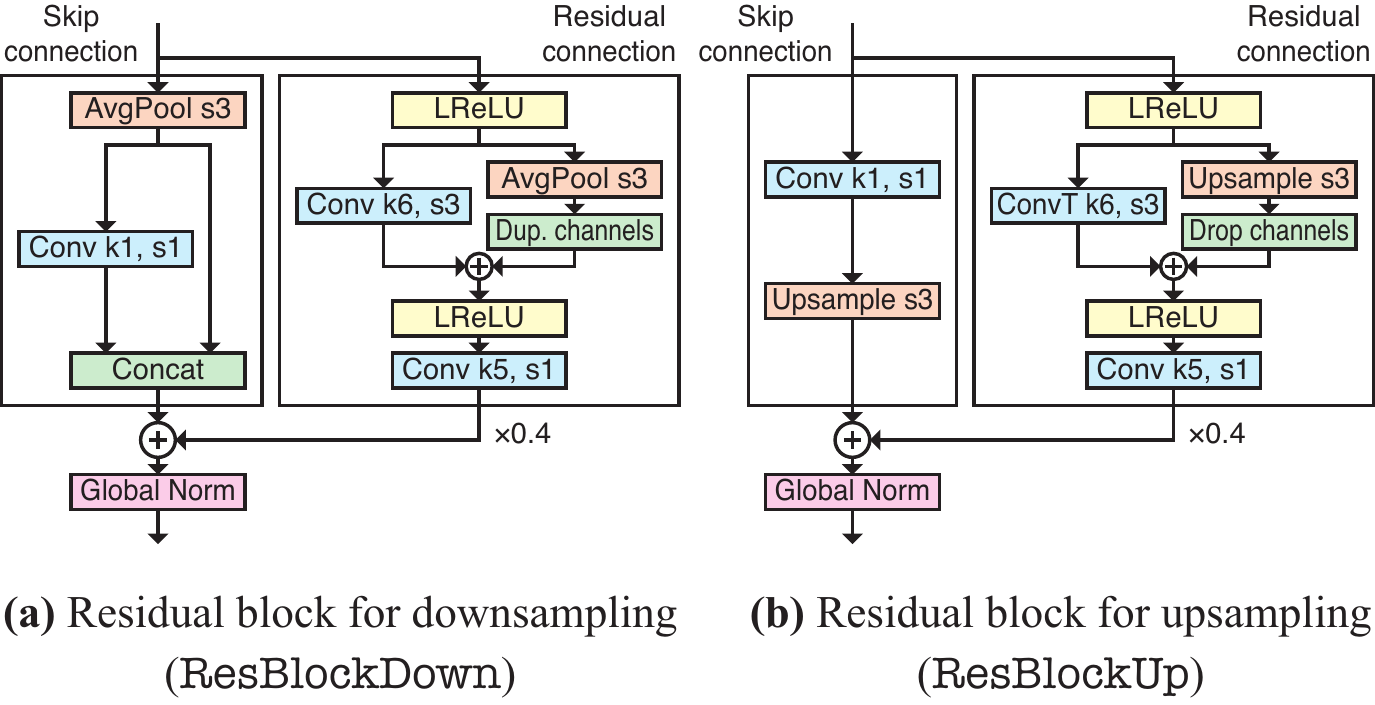}}
  \caption{Detailed architectures of residual blocks in a Wave-U-Net discriminator. \texttt{k}$x$ and \texttt{s}$y$ indicate the kernel size of $x$ and stride of $y$, respectively. The number of channels increases and decreases in the first convolution layer (\texttt{Conv}) and the first transposed convolution layer (\texttt{ConvT}) for downsampling and upsampling, respectively. A leaky rectified linear unit (\texttt{LReLU})~\cite{AMaasICML2013} with a negative slope of $0.1$ is used as an activation function. ``Dup. channels'' is the abbreviation for ``Duplicate channels'', ``+'' denotes the addition of features, and ``$\times 0.4$'' indicates that the features are scaled by a constant of $0.4$.}
  \label{fig:resblocks}
\end{figure}

\subsection{Techniques for stabilizing GAN training}
\label{subsec:waveunetd_techniques}

It is known that GANs are sensitive to architectural design because they conduct an unstable min-max game during training. This also holds true in this case, and preliminary experiments indicated that the direct use of a conventional Wave-U-Net~\cite{SPascualIS2017,DStollerISMIR2018} architecture saturated adversarial losses. This phenomenon could occur when a discriminator restricts itself to specific features that make it possible to distinguish real from fake data easily and experiences difficulties in propagating adequate information to a generator. Through extensive exploration, it was determined that careful normalization and the introduction of residual connections are necessary and sufficient to alleviate this deterioration.

\smallskip\noindent\textbf{Normalization.}
A previous study~\cite{KKumarNeurIPS2019} showed that in speech synthesis, a discriminator is sensitive to normalization, and normalization commonly used in image synthesis, such as spectral normalization~\cite{TMiyatoICLR2018} and instance normalization~\cite{DUlyanovArXiv2016}, does not work well. Alternatively, a previous study~\cite{KKumarNeurIPS2019} applied weight normalization~\cite{TSalimansNIPS2016}, which performed well with the conventional discriminator. However, this study determined that weight normalization is not sufficient to stabilize training with the Wave-U-Net discriminator, possibly because the Wave-U-Net discriminator, which consists of an encoder and a decoder, is deeper than typical discriminators, which include only an encoder, and it is difficult to avoid restricting itself to specific features. To alleviate this problem, this study proposes global normalization (\texttt{Global Norm}), which is defined as
\begin{flalign}
  \label{eq:global_norm}
  b = a \left/ \sqrt{\frac{1}{N}\sum\nolimits_{i=1}^N (a^i)^2 + \epsilon} \right. ,
\end{flalign}
where $\epsilon = 10^{-8}$; $N$ denotes the total number of features; $a$ and $b$ represent the original and normalized feature vectors, respectively; and $a^i$ indicates the $i$-th feature in $a$. This normalization prevents the discriminator from restricting itself to specific features by regularizing the norms of the features within a constant range.

One possible alternative is layer normalization~\cite{JBaArXiv2016}. Unlike global normalization, layer normalization uses trainable bias and gain parameters and conducts a sample-wise zero-mean shift. Preliminary experiments indicated that the first property invoked training instability and the second deteriorated cFW2VD~\cite{TKanekoICASSP2022}, which is a metric for assessing speech quality and is explained in detail in Section~\ref{subsec:eval_vocoder}. Based on these preliminary experiments, this study used global normalization, which only scales features without any trainable parameters.

\smallskip\noindent\textbf{Residual connections.}
This study empirically determined global normalization to be effective even for a non-residual Wave-U-Net discriminator. However, adversarial losses still approached saturation at the end of the training, possibly because the Wave-U-Net discriminator was deeper, as previously discussed. Hence, this study introduces residual connections~\cite{KHeCVPR2016} to prevent the vanishing gradient problem that typically occurs in deep networks. Inspired by the success of a previous GAN~\cite{LMeschederICML2018}, the output of the residual connection was scaled by a constant of $0.4$. This technique contributed slightly to stabilizing the training. Figure~\ref{fig:resblocks} shows the detailed architecture of the residual blocks equipped with the aforementioned modules.

\section{Experiments}
\label{sec:experiments}

Experiments were conducted to examine the general validity of the proposed ideas, that is, whether a Wave-U-Net discriminator can be used as an alternative to a typical ensemble of discriminators in various situations. This study first examined the dataset independence by applying the presented ideas to neural vocoders (particularly HiFI-GAN~\cite{JKongNeurIPS2020}) on various datasets. Subsequently, task independence was investigated by applying the presented ideas to an end-to-end TTS system (particularly VITS~\cite{JKimICML2021}). The details are discussed in Sections~\ref{subsec:eval_vocoder} and \ref{subsec:eval_end_to_end}. Audio samples are available from the link on the first page.\footnoteref{foot:samples}

\subsection{Evaluation on neural vocoders}
\label{subsec:eval_vocoder}

\textbf{Implementation.}
We first examined the effectiveness of a Wave-U-Net discriminator (hereafter denoted by \textit{Wave-U-Net $D$} for simplicity) when used to train a neural vocoder. Specifically, \textit{HiFI-GAN}~\cite{JKongNeurIPS2020} was used as the baseline and the difference in performance was examined when a Wave-U-Net discriminator was used as an alternative to the original ensemble of discriminators.
The proposed model was implemented based on the official implementation of HiFi-GAN,\footnote{\url{https://github.com/jik876/hifi-gan}} and the default training settings were used. The networks were trained using the AdamW optimizer~\cite{ILoshchilovICLR2019} with $\beta_1 = 0.8$, $\beta_2 = 0.99$, and weight decay $\lambda = 0.01$. The learning rate decayed exponentially by $0.999$ at fixed intervals with an initial learning rate of $2 \times 10^{-4}$. The batch size was set to $16$ with a segment length of $8192$.

\smallskip\noindent\textbf{Datasets.}
To examine the dataset independence, the models were evaluated based on three different datasets: \textit{LJSpeech}~\cite{ljspeech17}, which consists of 13,100 audio clips of a single English female speaker, and 12,500, 100, and 500 audio clips were used for training, validation, and evaluation, respectively. Specifically, data splitting was conducted based on an open-source code.\footnoteref{foot:vits} \textit{VCTK}~\cite{JYamagishiCSTR2016}, which includes 44,081 audio clips from 108 different English speakers, and 41,921, 1,080, and 1,080 audio clips were used for training, validation, and evaluation, respectively. The audio clips were divided based on an open-source code.\footnote{\label{foot:pwg}\url{https://github.com/kan-bayashi/ParallelWaveGAN}} \textit{JSUT}~\cite{RSonobeArXiv2017}, which is composed of 7,696 audio clips from a single Japanese female speaker, and 7,196, 250, and 250 audio clips were used for training, validation, and evaluation, respectively. Specifically, the audio clips were split according to an open-source code.\footnoteref{foot:pwg} Following the study on HiFi-GAN~\cite{JKongNeurIPS2020}, audio clips were sampled at $22.05$ kHz, and $80$-dimensional log-mel spectrograms with an FFT size of $1024$, hop length of $256$, and window length of $1024$ were extracted from the audio clips. Neural vocoders were individually trained for each dataset.

\begin{table}[t]
  \caption{Comparison of MOS with 95\% confidence intervals, cFW2VD, processing time, and number of parameters among neural vocoders on LJSpeech.}
  \label{tab:result_vocoder_ljspeech}
  \vspace{1mm}
  \newcommand{\spm}[1]{{\tiny$\pm$#1}}
  \setlength{\tabcolsep}{5pt}
  \renewcommand{\arraystretch}{0.95}
  \centering
  \scriptsize{
  \begin{tabularx}{\columnwidth}{lcccc}
    \toprule
    \multicolumn{1}{c}{\textbf{Model}} & \textbf{MOS}$\uparrow$ & \textbf{cFW2VD}$\downarrow$ & \textbf{Time}$\downarrow$ (s/batch) & \textbf{\# Param}$\downarrow$ (M)
    \\ \midrule
    Ground truth
    & 4.72\spm{0.07} & 0.000 & -- & --
    \\ \midrule
    HiFi-GAN
    & 4.57\spm{0.09} & 0.014 & 0.028 & 70.7
    \\
    w/ Wave-U-Net $D$
    & 4.54\spm{0.09} & 0.014 & 0.012 & \:\:4.9
    \\ \midrule
    MelGAN
    & 3.26\spm{0.14} & 0.193 & -- & --
    \\ \bottomrule    
  \end{tabularx}
  }
\end{table}

\begin{table}[t]
  \caption{Comparison of MOS with 95\% confidence intervals, cFW2VD, processing time, and number of parameters among neural vocoders on VCTK.}
  \label{tab:result_vocoder_vctk}
  \vspace{1mm}
  \newcommand{\spm}[1]{{\tiny$\pm$#1}}
  \setlength{\tabcolsep}{5pt}
  \renewcommand{\arraystretch}{0.95}
  \centering
  \scriptsize{
  \begin{tabularx}{\columnwidth}{lcccc}
    \toprule
    \multicolumn{1}{c}{\textbf{Model}} & \textbf{MOS}$\uparrow$ & \textbf{cFW2VD}$\downarrow$ & \textbf{Time}$\downarrow$ (s/batch) & \textbf{\# Param}$\downarrow$ (M)
    \\ \midrule
    Ground truth
    & 4.62\spm{0.07} & 0.000 & -- & --
    \\ \midrule
    HiFi-GAN
    & 4.37\spm{0.10} & 0.040 & 0.028 & 70.7
    \\
    w/ Wave-U-Net $D$
    & 4.44\spm{0.09} & 0.041 & 0.012 & \:\:4.9
    \\ \midrule
    MelGAN
    & 3.60\spm{0.13} & 0.301 & -- & --
    \\ \bottomrule    
  \end{tabularx}
  }
\end{table}

\begin{table}[t]
  \caption{Comparison of MOS with 95\% confidence intervals, cFW2VD, processing time, and number of parameters on among neural vocoders on JSUT.}
  \label{tab:result_vocoder_jsut}
  \vspace{1mm}
  \newcommand{\spm}[1]{{\tiny$\pm$#1}}
  \setlength{\tabcolsep}{5pt}
  \renewcommand{\arraystretch}{0.95}
  \centering
  \scriptsize{
  \begin{tabularx}{\columnwidth}{lcccc}
    \toprule
    \multicolumn{1}{c}{\textbf{Model}} & \textbf{MOS}$\uparrow$ & \textbf{cFW2VD}$\downarrow$ & \textbf{Time}$\downarrow$ (s/batch) & \textbf{\# Param}$\downarrow$ (M)
    \\ \midrule
    Ground truth
    & 4.75\spm{0.07} & 0.000 & -- & --
    \\ \midrule
    HiFi-GAN
    & 4.67\spm{0.07} & 0.039 & 0.028 & 70.7
    \\
    w/ Wave-U-Net $D$
    & 4.69\spm{0.07} & 0.038 & 0.012 & \:\:4.9
    \\ \midrule
    MelGAN
    & 3.68\spm{0.13} & 0.222 & -- & --
    \\ \bottomrule    
  \end{tabularx}
  }
\end{table}

\smallskip\noindent\textbf{Evaluation metrics.}
We employed four metrics to evaluate the performance: Mean opinion score (\textit{MOS}) tests were conducted to evaluate the perceptual quality. Twenty utterances were randomly selected from the evaluation set, and the mel spectrograms extracted from the utterances were used as the vocoder input. The \textit{ground-truth} speech and speech synthesized by \textit{MelGAN}~\cite{KKumarNeurIPS2019} were used as anchor samples. These tests were conducted online with 11 listeners participating. The listeners were asked to judge the speech quality using a five-point scale: 1 = bad, 2 = poor, 3 = fair, 4 = good, and 5 = excellent. The conditional Fr\'{e}chet wav2vec distance (\textit{cFW2VD})~\cite{TKanekoICASSP2022} was used as an objective metric for assessing the speech quality. This metric measures the distance between the real and generative distributions in a wav2vec 2.0~\cite{ABaevskiNeurIPS2020} feature space conditioned on the text. The smaller the value, the higher is the similarity between the real and synthesized speech. For the training speed, the \textit{time} required for a discriminator to forward-propagate real and synthesized speech in a batch was measured on a single NVIDIA A100 GPU. Here, the smaller the value, the faster the speed. Number of parameters (\textit{\#~Param}) was used to investigate the model size. The smaller the value, the more lightweight the model.

\smallskip\noindent\textbf{Results.}
Tables~\ref{tab:result_vocoder_ljspeech}, \ref{tab:result_vocoder_vctk}, and \ref{tab:result_vocoder_jsut} present the results for LJSpeech, VCTK, and JSUT, respectively. It was determined that, across all the datasets, the proposed model (\textit{w/ Wave-U-Net $D$}) achieved comparable speech quality with the baseline (\textit{HiFi-GAN}) in terms of MOS\footnote{The $p$-values for the results in Tables~\ref{tab:result_vocoder_ljspeech}, \ref{tab:result_vocoder_vctk}, and \ref{tab:result_vocoder_jsut} were $0.622$, $0.533$, and $0.708$, respectively, in the Mann-Whitney U tests. The results indicate that the two models do not differ significantly regarding the $p$-value $< 0.05$.} and cFW2VD with a 2.31 times faster and 14.5 times more lightweight discriminator. These results indicate that the Wave-U-Net discriminator can be used as an alternative to the HiFi-GAN discriminator regardless of the dataset.

\begin{table}[t]
  \caption{Comparison of MOS with 95\% confidence intervals, cFW2VD, processing time, and number of parameters among TTS systems.}
  \label{tab:result_end_to_end}
  \vspace{1mm}
  \newcommand{\spm}[1]{{\tiny$\pm$#1}}
  \setlength{\tabcolsep}{3pt}
  \renewcommand{\arraystretch}{0.95}
  \centering
  \scriptsize{
  \begin{tabularx}{\columnwidth}{lcccc}
    \toprule
    \multicolumn{1}{c}{\textbf{Model}} & \textbf{MOS}$\uparrow$ & \textbf{cFW2VD}$\downarrow$ & \textbf{Time}$\downarrow$ (s/batch) & \textbf{\# Param}$\downarrow$ (M)
    \\ \midrule
    Ground truth
    & 4.65\spm{0.07} & 0.000 & -- & --
    \\ \midrule
    VITS
    & 4.51\spm{0.09} & 0.109 & 0.030 & 46.7
    \\
    w/ Wave-U-Net $D$
    & 4.54\spm{0.09} & 0.101 & 0.016 & \:\:4.9
    \\ \midrule
    Tacotron 2 + HiFi-GAN
    & 3.78\spm{0.14} & 0.261 & -- & --
    \\ \bottomrule    
  \end{tabularx}
  }
\end{table}

\subsection{Evaluation on end-to-end TTS}
\label{subsec:eval_end_to_end}

\textbf{Implementation.}
To investigate task independence, the proposed ideas were applied to end-to-end TTS. Specifically, \textit{VITS}~\cite{JKimICML2021} was used as the baseline, and the performance difference was examined when the original ensemble of discriminators was replaced with a Wave-U-Net discriminator. The proposed model was implemented based on the official implementation of VITS\footnote{\label{foot:vits}\url{https://github.com/jaywalnut310/vits}} and the default training settings were used. The networks were trained using the AdamW optimizer~\cite{ILoshchilovICLR2019} with $\beta_1 = 0.8$, $\beta_2 = 0.99$, and weight decay $\lambda = 0.01$. The learning rate decayed exponentially by $0.999^{\frac{1}{8}}$ in fixed intervals with an initial learning rate of $2 \times 10^{-4}$. The batch size was set to $64$ with a segment length of $8192$.

\smallskip\noindent\textbf{Dataset.}
LJSpeech~\cite{ljspeech17} was used for the experiment. The data splitting setting is the same as that used in the experiment described in Section~\ref{subsec:eval_vocoder}.

\smallskip\noindent\textbf{Evaluation metrics.}
The same evaluation metrics are used to evaluate the neural vocoders (Section~\ref{subsec:eval_vocoder}). In the MOS test, the anchor samples were altered because of the task differences. In this experiment, \textit{ground-truth} speech and speech synthesized using a combination of Tacotron 2~\cite{JShenICASSP2018} and HiFi-GAN~\cite{JKongNeurIPS2020} (\textit{Tacotron 2 + HiFi-GAN}) were used as anchor samples.

\smallskip\noindent\textbf{Results.}
Table~\ref{tab:result_end_to_end} summarizes the results. The results show that the proposed model (\textit{w/ Wave-U-Net $D$}) achieved speech quality comparable to that of the baseline model (\textit{VITS}) in terms of the MOS\footnote{The $p$-value was $0.345$ in the Mann-Whitney U test. This score indicates that the two models do not differ significantly regarding the $p$-value $< 0.05$.} and cFW2VD with a 1.90 times faster and 9.62 times more lightweight discriminator. These results verified the task independence of the Wave-U-Net discriminator.

\section{Conclusion}
\label{sec:conclusion}

This study proposes a Wave-U-Net discriminator, which is a single but expressive discriminator that assesses a waveform in a sample-wise manner with the same resolution as the input signal while extracting multilevel features via an encoder and decoder with skip connections. The experimental results demonstrate that a Wave-U-Net discriminator can be used as an alternative to a typical ensemble of discriminators while maintaining speech quality, reducing the model size, and accelerating the training speed. Although the general utility of a Wave-U-Net discriminator has been demonstrated to some extent, there are several other tasks that are beyond the scope of this study, such as singing speech synthesis, emotional speech synthesis, and music synthesis. Utilization of the proposed ideas is a valuable future research topic.

\smallskip\noindent\textbf{Acknowledgements.}
This work was supported by JST CREST Grant Number JPMJCR19A3, Japan.

\vfill\pagebreak

\bibliographystyle{IEEEbib}
\renewcommand{\baselinestretch}{0.92}
{\footnotesize\bibliography{strings,refs}}

\begin{thebibliography}{10}

\bibitem{AOordArXiv2016}
A\"aron van~den Oord, Sander Dieleman, Heiga Zen, Karen Simonyan, Oriol
  Vinyals, Alex Graves, Nal Kalchbrenner, Andrew Senior, and Koray Kavukcuoglu,
\newblock ``{WaveNet}: A generative model for raw audio,''
\newblock {\em arXiv preprint arXiv:1609.03499}, 2016.

\bibitem{NKalchbrennerICML2018}
Nal Kalchbrenner, Erich Elsen, Karen Simonyan, Seb Noury, Norman Casagrande,
  Edward Lockhart, Florian Stimberg, A\"aron van~den Oord, Sander Dieleman, and
  Koray Kavukcuoglu,
\newblock ``Efficient neural audio synthesis,''
\newblock in {\em Proc. ICML}, 2018, pp. 2410--2419.

\bibitem{AOordICML2018}
A\"aron van~den Oord, Yazhe Li, Igor Babuschkin, Karen Simonyan, Oriol Vinyals,
  Koray Kavukcuoglu, George van~den Driessche, Edward Lockhart, Luis Cobo,
  Florian Stimberg, Norman Casagrande, Dominik Grewe, Seb Noury, Sander
  Dieleman, Erich Elsen, Nal Kalchbrenner, Heiga Zen, Alex Graves, Helen King,
  Tom Walters, Dan Belov, and Demis Hassabis,
\newblock ``Parallel {W}ave{N}et: Fast high-fidelity speech synthesis,''
\newblock in {\em Proc. ICML}, 2018, pp. 3918--3926.

\bibitem{RPrengerICASSP2019}
Ryan Prenger, Rafael Valle, and Bryan Catanzaro,
\newblock ``{WaveGlow}: A flow-based generative network for speech synthesis,''
\newblock in {\em Proc. ICASSP}, 2019, pp. 3617--3621.

\bibitem{KKumarNeurIPS2019}
Kundan Kumar, Rithesh Kumar, Thibault de~Boissiere, Lucas Gestin, Wei~Zhen
  Teoh, Jose Sotelo, Alexandre de~Br{\'e}bisson, Yoshua Bengio, and Aaron
  Courville,
\newblock ``{MelGAN}: Generative adversarial networks for conditional waveform
  synthesis,''
\newblock in {\em Proc. NeurIPS}, 2019, pp. 14910--14921.

\bibitem{RYamamotoICASSP2020}
Ryuichi Yamamoto, Eunwoo Song, and Jae-Min Kim,
\newblock ``{Parallel WaveGAN}: A fast waveform generation model based on
  generative adversarial networks with multi-resolution spectrogram,''
\newblock in {\em Proc. ICASSP}, 2020, pp. 6199--6203.

\bibitem{JYangIS2020}
Jinhyeok Yang, Junmo Lee, Youngik Kim, Hoonyoung Cho, and Injung Kim,
\newblock ``{VocGAN}: A high-fidelity real-time vocoder with a
  hierarchically-nested adversarial network,''
\newblock in {\em Proc. Interspeech}, 2020, pp. 200--204.

\bibitem{JKongNeurIPS2020}
Jungil Kong, Jaehyeon Kim, and Jaekyoung Bae,
\newblock ``{HiFi-GAN}: Generative adversarial networks for efficient and high
  fidelity speech synthesis,''
\newblock in {\em Proc. NeurIPS}, 2020, pp. 17022--17033.

\bibitem{GYangSLT2021}
Geng Yang, Shan Yang, Kai Liu, Peng Fang, Wei Chen, and Lei Xie,
\newblock ``{Multi-band MelGAN}: Faster waveform generation for high-quality
  text-to-speech,''
\newblock in {\em Proc. SLT}, 2021, pp. 492--498.

\bibitem{AMustafaICASSP2021}
Ahmed Mustafa, Nicola Pia, and Guillaume Fuchs,
\newblock ``{StyleMelGAN}: An efficient high-fidelity adversarial vocoder with
  temporal adaptive normalization,''
\newblock in {\em Proc. ICASSP}, 2021, pp. 6034--6038.

\bibitem{TKanekoICASSP2022}
Takuhiro Kaneko, Kou Tanaka, Hirokazu Kameoka, and Shogo Seki,
\newblock ``{iSTFTNet}: Fast and lightweight mel-spectrogram vocoder
  incorporating inverse short-time {Fourier} transform,''
\newblock in {\em Proc. ICASSP}, 2022, pp. 6207--6211.

\bibitem{TKanekoIS2022}
Takuhiro Kaneko, Hirokazu Kameoka, Kou Tanaka, and Shogo Seki,
\newblock ``{MISRNet}: Lightweight neural vocoder using multi-input single
  shared residual blocks,''
\newblock in {\em Proc. Interspeech}, 2022, pp. 1631--1635.

\bibitem{NChenICLR2021}
Nanxin Chen, Yu~Zhang, Heiga Zen, Ron~J. Weiss, Mohammad Norouzi, and William
  Chan,
\newblock ``{WaveGrad}: Estimating gradients for waveform generation,''
\newblock in {\em Proc. ICLR}, 2021.

\bibitem{ZKongICLR2021}
Zhifeng Kong, Wei Ping, Jiaji Huang, Kexin Zhao, and Bryan Catanzaro,
\newblock ``{DiffWave}: A versatile diffusion model for audio synthesis,''
\newblock in {\em Proc. ICLR}, 2021.

\bibitem{KQianICML2019}
Kaizhi Qian, Yang Zhang, Shiyu Chang, Xuesong Yang, and Mark Hasegawa-Johnson,
\newblock ``{Auto-VC}: Zero-shot voice style transfer with only autoencoder
  loss,''
\newblock in {\em Proc. ICML}, 2019, pp. 5210--5219.

\bibitem{TKanekoIS2020}
Takuhiro Kaneko, Hirokazu Kameoka, Kou Tanaka, and Nobukatsu Hojo,
\newblock ``{CycleGAN-VC3}: Examining and improving {CycleGAN-VCs} for
  mel-spectrogram conversion,''
\newblock in {\em Proc. Interspeech}, 2020, pp. 2017--2021.

\bibitem{TKanekoICASSP2021}
Takuhiro Kaneko, Hirokazu Kameoka, Kou Tanaka, and Nobukatsu Hojo,
\newblock ``{MaskCycleGAN-VC}: Learning non-parallel voice conversion with
  filling in frames,''
\newblock in {\em Proc. ICASSP}, 2021, pp. 5919--5923.

\bibitem{WPingICLR2019}
Wei Ping, Kainan Peng, and Jitong Chen,
\newblock ``{ClariNet}: Parallel wave generation in end-to-end
  text-to-speech,''
\newblock in {\em Proc. ICLR}, 2019.

\bibitem{YRenICLR2021}
Yi~Ren, Chenxu Hu, Xu~Tan, Tao Qin, Sheng Zhao, Zhou Zhao, and Tie-Yan Liu,
\newblock ``{FastSpeech 2}: Fast and high-quality end-to-end text to speech,''
\newblock in {\em Proc. ICLR}, 2021.

\bibitem{JDonahueICLR2021}
Jeff Donahue, Sander Dieleman, Miko{\l}aj Bi{\'n}kowski, Erich Elsen, and Karen
  Simonyan,
\newblock ``End-to-end adversarial text-to-speech,''
\newblock in {\em Proc. ICLR}, 2021.

\bibitem{JKimICML2021}
Jaehyeon Kim, Jungil Kong, and Juhee Son,
\newblock ``Conditional variational autoencoder with adversarial learning for
  end-to-end text-to-speech,''
\newblock in {\em Proc. ICML}, 2021, pp. 5530--5540.

\bibitem{IGoodfellowNIPS2014}
Ian~J. Goodfellow, Jean Pouget-Abadie, Mehdi Mirza, Bing Xu, David
  Warde-Farley, Sherjil Ozair, Aaron Courville, and Yoshua Bengio,
\newblock ``Generative adversarial nets,''
\newblock in {\em Proc. NIPS}, 2014, pp. 2672--2680.

\bibitem{SPascualIS2017}
Santiago Pascual, Antonio Bonafonte, and Joan Serra,
\newblock ``{SEGAN}: Speech enhancement generative adversarial network,''
\newblock in {\em Proc. Interspeech}, 2017, pp. 3642--3646.

\bibitem{DStollerISMIR2018}
Daniel Stoller, Sebastian Ewert, and Simon Dixon,
\newblock ``{Wave-U-Net}: A multi-scale neural network for end-to-end audio
  source separation,''
\newblock in {\em Proc. ISMIR}, 2018, pp. 334--340.

\bibitem{ESchonfeldCVPR2020}
Edgar Schonfeld, Bernt Schiele, and Anna Khoreva,
\newblock ``A {U-Net} based discriminator for generative adversarial
  networks,''
\newblock in {\em Proc. CVPR}, 2020, pp. 8207--8216.

\bibitem{ORonnebergerMICCAI2015}
Olaf Ronneberger, Philipp Fischer, and Thomas Brox,
\newblock ``{U-Net}: Convolutional networks for biomedical image
  segmentation,''
\newblock in {\em Proc. MICCAI}, 2015, pp. 234--241.

\bibitem{ljspeech17}
Keith Ito and Linda Johnson,
\newblock ``The {LJ} speech dataset,''
  \url{https://keithito.com/LJ-Speech-Dataset/}, 2017.

\bibitem{JYamagishiCSTR2016}
Junichi Yamagishi, Christophe Veaux, and Kirsten MacDonald,
\newblock ``{CSTR VCTK} corpus: {English} multi-speaker corpus for {CSTR} voice
  cloning toolkit,''
\newblock {\em The Centre for Speech Technology Research}, 2016.

\bibitem{RSonobeArXiv2017}
Ryosuke Sonobe, Shinnosuke Takamichi, and Hiroshi Saruwatari,
\newblock ``{JSUT corpus}: free large-scale {Japanese} speech corpus for
  end-to-end speech synthesis,''
\newblock {\em arXiv preprint arXiv:1711.00354}, 2017.

\bibitem{XMaoICCV2017}
Xudong Mao, Qing Li, Haoran Xie, Raymond~Y.K. Lau, Zhen Wang, and Stephen~Paul
  Smolley,
\newblock ``Least squares generative adversarial networks,''
\newblock in {\em Proc. ICCV}, 2017, pp. 2794--2802.

\bibitem{ALarsenICML2016}
Anders Boesen~Lindbo Larsen, S{\o}ren~Kaae S{\o}nderby, Hugo Larochelle, and
  Ole Winther,
\newblock ``Autoencoding beyond pixels using a learned similarity metric,''
\newblock in {\em Proc. ICML}, 2016, pp. 1558--1566.

\bibitem{TKanekoIS2017}
Takuhiro Kaneko, Hirokazu Kameoka, Kaoru Hiramatsu, and Kunio Kashino,
\newblock ``Sequence-to-sequence voice conversion with similarity metric
  learned using generative adversarial networks,''
\newblock in {\em Proc. Interspeech}, 2017, pp. 1283--1287.

\bibitem{AMaasICML2013}
Andrew~L. Maas, Awni~Y. Hannun, and Andrew~Y. Ng,
\newblock ``Rectifier nonlinearities improve neural network acoustic models,''
\newblock in {\em Proc. ICML}, 2013.

\bibitem{TMiyatoICLR2018}
Takeru Miyato, Toshiki Kataoka, Masanori Koyama, and Yuichi Yoshida,
\newblock ``Spectral normalization for generative adversarial networks,''
\newblock in {\em Proc. ICLR}, 2018.

\bibitem{DUlyanovArXiv2016}
Dmitry Ulyanov, Andrea Vedaldi, and Victor Lempitsky,
\newblock ``Instance normalization: The missing ingredient for fast
  stylization,''
\newblock {\em arXiv preprint arXiv:1607.08022}, 2016.

\bibitem{TSalimansNIPS2016}
Tim Salimans and Diederik~P. Kingma,
\newblock ``Weight normalization: A simple reparameterization to accelerate
  training of deep neural networks,''
\newblock in {\em Proc. NIPS}, 2016, pp. 901--909.

\bibitem{JBaArXiv2016}
Jimmy~Lei Ba, Jamie~Ryan Kiros, and Geoffrey~E. Hinton,
\newblock ``Layer normalization,''
\newblock {\em arXiv preprint arXiv:1607.06450}, 2016.

\bibitem{KHeCVPR2016}
Kaiming He, Xiangyu Zhang, Shaoqing Ren, and Jian Sun,
\newblock ``Deep residual learning for image recognition,''
\newblock in {\em Proc. CVPR}, 2016, pp. 770--778.

\bibitem{LMeschederICML2018}
Lars Mescheder, Andreas Geiger, and Sebastian Nowozin,
\newblock ``Which training methods for {GANs} do actually converge?,''
\newblock in {\em Proc. ICML}, 2018, pp. 3481--3490.

\bibitem{ILoshchilovICLR2019}
Ilya Loshchilov and Frank Hutter,
\newblock ``Decoupled weight decay regularization,''
\newblock in {\em Proc. ICLR}, 2019.

\bibitem{ABaevskiNeurIPS2020}
Alexei Baevski, Henry Zhou, Abdelrahman Mohamed, and Michael Auli,
\newblock ``wav2vec 2.0: A framework for self-supervised learning of speech
  representations,''
\newblock in {\em Proc. NeurIPS}, 2020, pp. 12449--12460.

\bibitem{JShenICASSP2018}
Jonathan Shen, Ruoming Pang, Ron~J. Weiss, Mike Schuster, Navdeep Jaitly,
  Zongheng Yang, Zhifeng Chen, Yu~Zhang, Yuxuan Wang, RJ~Skerrv-Ryan, Rif~A.
  Saurous, Yannis Agiomyrgiannakis, and Yonghui Wu,
\newblock ``Natural {TTS} synthesis by conditioning {WaveNet} on mel
  spectrogram predictions,''
\newblock in {\em Proc. ICASSP}, 2018, pp. 4779--4783.

\end{thebibliography}

\end{document}